\documentclass{elsart}
\usepackage{amsfonts}
\usepackage{amsmath}
\usepackage{amssymb}
\usepackage{graphicx}

\begin{document}

\begin{frontmatter}

\title{Creating a self-induced dark spontaneous-force optical trap for neutral atoms}

\collab{S. R. Muniz, K. M. F. Magalh\~{a}es, E. A. L. Henn, L. G. Marcassa and 
V. S. Bagnato.}

\address{Instituto de F\'{\i}sica de S\~{a}o Carlos, Universidade de S\~{a}o Paulo, Caixa Postal 
369,  S\~{a}o Carlos, SP  -  CEP 13560-970 - Brazil }

\begin{abstract}
This communication describes the observation of a new type of dark spontaneous-force
optical trap (dark SPOT) obtained  without the use of a mask blocking the central part 
of the repumper laser beam. We observe that loading a magneto-optical trap (MOT) 
from a continuous and intense flux of slowed atoms and by  appropriately tuning the 
frequency of the repumper laser is possible to achieve basically the same effect of the 
dark SPOT, using a simpler apparatus. This work characterizes the new system through 
measurements of absorption and fluorescence imaging of the atomic cloud and presents
a very simple model to explain the main features of our observations.  
We believe that this new approach may simplify the current experiments to produce 
quantum degenerated gases
\end{abstract}


\end{frontmatter}

\section*{Introduction}

Laser cooling and trapping techniques have been very useful in the studies
of atomic physics. Particularly the magneto-optical trap (MOT) has shown to
be a versatile method to produce cold and relatively dense atomic samples.
Its development is certainly one of the main responsible for the wide
spreading of the cold and ultracold matter studies and applications \cite%
{Nature}. In many of these studies is important to have a very dense sample
of ultracold atoms populating the lower hyperfine ground state. However, in
practice, there is a limit to the highest density attainable in a MOT, which
is typically on the order of $10^{10}$ $atoms/cm^{3}$.

As it is well known, the limit to the density in a MOT is consequence of two
major processes: cold-collision induced losses and secondary light
scattering. The first impediment involves the cold collision of ground and
excited state atoms within the trap, which allows transforming part of the
excitation energy of the photon into kinetic energy of the colliding pair.
This process eventually provides enough energy to the atoms to leave the
trap \cite{WBZ}. In this case, higher densities imply in a larger collisions
rate and leads to higher losses. Since the trap loss rate is usually on the
order of $10^{-11}cm^{3}/s$, one has in practice samples with atomic\
densities on the order of $10^{10}cm^{-3}$. Secondary light scattering
imposes another limitation to these systems. The reabsorption of emitted
photons causes an outward radiation pressure \cite{WSW} that prevents
further increase of the density. Due to this second mechanism the loading
normally occurs at constant density, where an increase in the number of
atoms results in a increase of the volume \cite{WFH}.

In order to overcome both limitations Ketterle et al \cite{KDJ} developed
the so called dark spontaneous-force optical trap, or simply \ `dark SPOT',
where the trapped atoms were spatially confined in a hyperfine ground state
which does not interact with the trapping laser frequencies, staying
therefore in the dark. This scheme provides an effective way to supplant the
light scattering that causes the limitations mentioned before. This was
accomplished by placing a physical obstacle to block the central region of
the repumper laser beams. Atoms in the dark region are rapidly pumped to the
lower energy ground state and do not interact with the lasers. The absence
of light scattering in inner and most dense portion of the cloud causes a
tremendous decrease in the repulsive force induced by photon secondary
scattering as well as the ground-excited collisions rates and allowed them
to obtain approximately $5\times 10^{10}cm^{-3}$\ sodium atoms in this type
of "dark MOT". One of the main feature of the dark SPOT is to produce a very
dense and compact distribution of ground state atoms, located in the center
of the atomic trapped cloud. This achievement was an important step towards
magnetic trapping of these atoms and the further realization of Bose
Einstein Condensation \cite{AEM}.

In this paper we demonstrate a similar trap without the use of any obstacle
in the repumper beam. The dark region, in our case, is created by the own
atoms absorption. Using a continuous and intense flux of slowed sodium atoms
in their $3S_{1/2}(F=1)$ ground state and adjusting the intensity and
frequency of the repumper laser, we were able to capture about $10^{9}$
atoms in a configuration where the repumper beam is severally absorbed by
the outer part of the atomic cloud, in this way the atoms in the center of
the trap stay in a\ "dark ground state". We named this situation a
self-induced dark MOT (SDMOT). It represents an improvement in the
achievement of high density because it can be done even when a single
electro-optically modulated laser beam is used for trapping the atoms. In
the next sections we present a description of our experimental apparatus,
followed by the results obtained by absorption of a probe laser beam.
Finally we also present a simple model in order to explain some of the main
features of our system.

\section{Experimental Setup}

A schematic diagram of our experimental setup is showed in figure 1. An
effusive sodium beam is decelerated in a tapered solenoid by the
Zeeman-tuned technique \cite{BSM}. After the solenoid there is an extra coil
that allows to extract an intense flux of slowed atoms \cite{FFZ}. These
atoms are already in the lower hyperfine ground state, as discussed in the
next section of this paper and explained in more details in reference \cite%
{NZB} . \ In this configuration, to avoid any zero crossing in the value of
the magnetic field, the current circulation in the extra coil is such that
its field lines smoothly match the field lines of the MOT coils.

\begin{figure}[tbh]
\begin{center}
\includegraphics[width=4.5in]{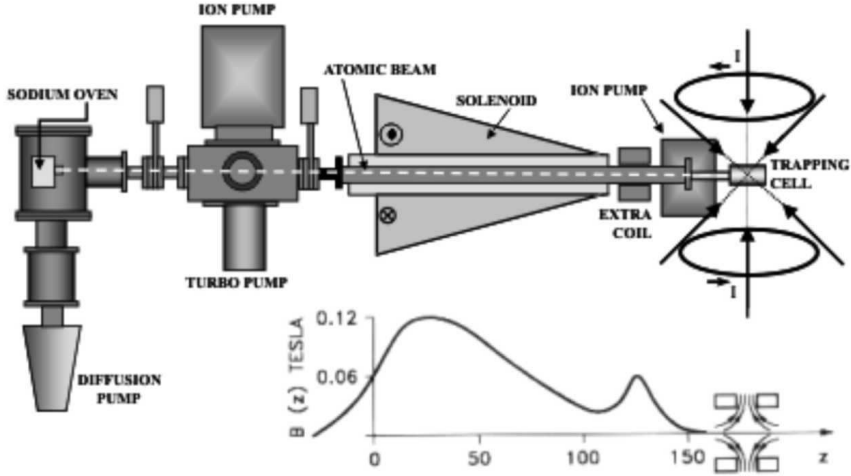}
\end{center}
\caption{ \emph{Schematic diagram of the experimental system. Atoms emerging
from an effusive oven are decelerated and trapped in a MOT aligned with
slowing tube. The magnetic field profile is presented in the figure detail,
where z in given in centimeters and B in tesla.} }
\end{figure}

Three ring-cavity dye lasers provide the light for the slowing, trapping and
repumping transitions. In order to do that, all the lasers are frequency
stabilized and peak-locked to the appropriate optical transition, using a
vapor cell and the saturated absorption signal. The laser frequencies are
easily tuned to the red of the sodium atomic transitions $%
3S_{1/2}(F=2)\longrightarrow 3P_{3/2}(F^{\prime }=3)$ and $%
3S_{1/2}(F=1)\longrightarrow 3P_{3/2}(F^{\prime }=2),$ at the specified $%
\Delta _{slower},$ $\Delta _{trap}$ and $\Delta _{repump}$ detunings,
through the use of acousto-optical modulators (AOM). Although in the
measurements described here we have used two independent lasers to produce
the SDMOT, we verified that similar results could be observed when a single
laser was used for trapping. In that case an electro-optical modulator (EOM)
tuned around 1.7 GHz is necessary to provide the repumping light. The main
reason for choosing the extra dye laser was to easily change the repumper
frequency, allowing for large repumping detunings. Once our home-built EOM
has a resonance cavity, to enhance the electric field applied to the
non-linear crystal \cite{EOMs}, it would be somewhat trickier to do the same
large frequency variation using only the EOM. In fact, the possibility of
easily changing the frequency of the repumper light might be one of the
reasons that allowed us to observe the above-mentioned effect.

The MOT is located in an ultra-high vacuum (UHV) glass cell, positioned at
the end of the deceleration tube and in-line with the slowing process, as
shown in figure 1. This configuration provides a very efficient coupling of
the slowed atoms into the MOT. To avoid the undesirable mechanical effect of
the strong slowing laser, tuned close to the $F=2\longrightarrow F^{\prime
}=3$\ transition, we use a variation of the Zeeman-tuned technique, which is
discussed in details in reference\ \cite{MMT}. The trap is created by six
independent laser beam tuned to the red of the trapping and repumping
transitions, respectively by the values $\Delta _{trap}$ and $\Delta
_{repump}.$ The repumper laser comes in four independent beams, aligned
collinearly with the trapping beam in the horizontal plane.

Once the MOT laser beams and the quadrupole coils are activated\ we capture
about $10^{9}$ atoms in the trap\ after the optimization of the MOT and
slowing process parameters. In our system, the atomic cloud can be
characterized either by fluorescence or absorption imaging using a
triggerable digital CCD camera.

\section{Emerging slowed atomic beam characteristics}

In order to characterize the outgoing flux of slow atoms we have used a
probe laser beam crossing the atomic beam at small angle. During this
characterization the MOT lasers and the quadrupole magnetic coils were
turned off. The fluorescence at the crossing position was imaged in a
photomultiplier tube and the analysis of the fluorescence, as function of
the probe frequency, allowed us to measure the velocity distribution of the
output flux after deceleration. This measurement also allowed us to
discriminate the population in each of the two ground states sub-levels.

Tuning the slower laser frequency close to the cycling transition $%
3S_{1/2}(F=2)\longrightarrow 3P_{3/2}(F^{\prime }=3)$ and with the slowing
magnet carrying a curent of about 45A, which produces a field approximately
given by $B(z)=1000\sqrt{1-z}\ +\ 100$, for B given in gauss and \textit{z}
in meters, one obtains a flux of slow atoms emerging in the $3S_{1/2}$ $(F=2)
$ ground state level with a peak velocity of about 200 m/s. To obtain lower
velocities at this condition the slowing laser frequency has to be tuned to
the blue of the transition. This type of behavior has been already
investigated by our group in the past \cite{NZB} and it is mainly due to the
adiabatic following of the atoms in the magnetic field during the slowing
process. The measured velocity distribution is presented in figure 2.
Decreasing the current of the slowing magnet to about 36A the situation is
completely different and a large flux of very slow atoms emerges from the
process in the $3S_{1/2}(F=1)$\ ground state level. The output velocity of
these atoms obey the relation $\ v_{out}$\ $\approx -\Delta _{slower}/k$, \
where $\Delta _{slower}$\ is the slower laser detuning and $k$\ is the wave
vector.

\begin{figure}[h]
\centering   \includegraphics[width=3in]{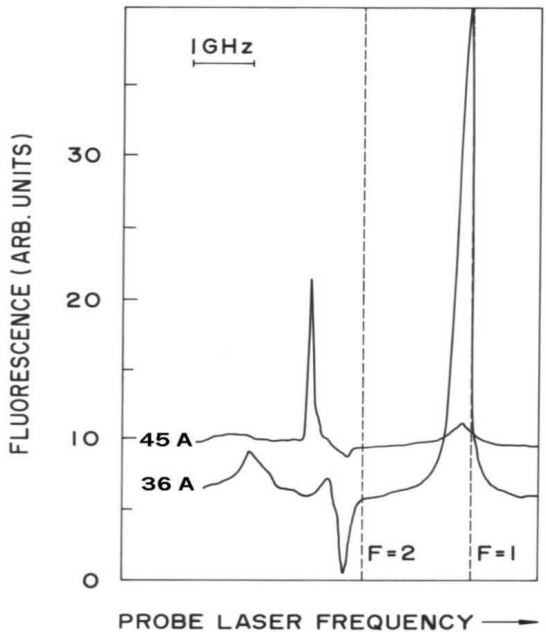}
\caption{ \emph{Fluorescence measurement of the outgoing distribution of
slowed atoms as function of a probe laser frequency. The measurements were
done at two different solenoid currents. At 45A the atomic flux is
predominantly at 3 S}$_{1/2}(F=2)$\emph{\ state, while at 36 A it is
predominantly 3 S}$_{1/2}(F=1)$\emph{\ atoms. Dashed lines show the position
of the atomic transitions.} }
\end{figure}

Although the atoms are decelerated in the cycling transition $%
3S_{1/2}(F=2)\longrightarrow 3P_{3/2}(F^{\prime }=3)$, as they approach the
end of the slowing solenoid, the lower amplitude of the field as well as the
configuration of the field lines create adequate conditions to optically
pump the atoms to the $3S_{1/2}(F=1)$ state. At this point the deceleration
process stops abruptly and the slowed atoms migrate out of the solenoid
without interacting with the slowing laser. We measured about $10^{10}$
atoms/sec emerging from the slowing process near zero velocity and already
in the $3S_{1/2}(F=1)$ state. The figure 2 also shows the velocity
distribution for the $3S_{1/2}(F=1)$ atoms when the second operating
condition is used. The peak velocity is nearly zero when $\Delta
_{slower}\thicksim 0$ and the distribution width is approximately 50 m/s.
Recalling a previous measurement of the capture velocity of a MOT \cite{SRM}%
, one sees that a considerable portion of this flux can be captured by the
trap.

\section{Trapping the decelerated atoms}

To be able to efficiently capture atoms in the MOT, the average atomic
velocity has to be on the order of the capture velocity of the trap. We use
a variant of the Zeeman-tuned technique \cite{MMT} to provide the slow atoms
for the MOT, which is kept in UHV conditions. Then, after activating the MOT
coils and lasers, as the slowed atoms emerging in $3S_{1/2}(F=1)$ get into
the capture region, they start to accumulate in the trap. The slowing laser
frequency and Zeeman magnet current are optimized for capturing the maximum
number atoms. In our experiment we used $\Delta _{trap}=-10MHz$, with
respect to the transition $3S_{1/2}(F=2)\longrightarrow 3P_{3/2}(F^{\prime
}=3)$, and $\Delta _{repump}=-30MHz$ with respect to the $%
3S_{1/2}(F=1)\longrightarrow 3P_{3/2}(F^{\prime }=0)$ transition. This
situation is represented in figure 3.\ The loading time was about 0.6
seconds and the total number of trapped atoms was around $10^{9}$atoms. Each
of the six beams of the trapping laser had about $30$ $mW$, and $8$ $mW$ on
each of the four repumping beams. The beams had a gaussian profile with
waist of approximately 1 cm FWHM.

\begin{figure}[h]
\centering   \includegraphics[width=3.5in]{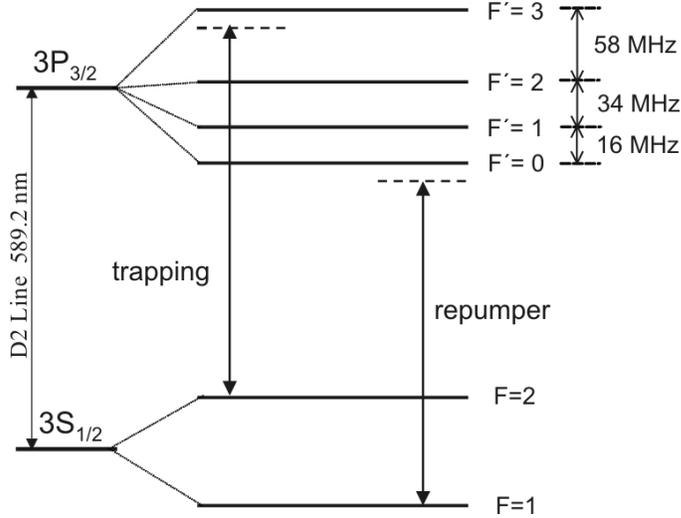}
\caption{ \emph{Diagram of sodium levels indicating the optical transitions
and the laser detunings involved.} }
\end{figure}

To understand how the process actually works, we can consider the capture
happening in the following way: atoms emerge from the slowing solenoid in
the $3S_{1/2}(F=1)$ ground state, interacting with the repumping\ and
trapping lasers they are pumped to $3S_{1/2}(F=2)$ state and initiate
cycling in the $3S_{1/2}(F=2)\longrightarrow 3P_{3/2}(F^{\prime }=3)$ strong
transition, where the spatially selective light pressure capture them in the
MOT . However, because of the high loading flux of atoms, there is a strong
absorption of the repumper laser across the trapped cloud. As consequence of
this fact, the repumper laser is attenuated and therefore atoms are no
longer efficiently repumped to $3S_{1/2}(F=2)$ state. The result is an
atomic cloud composed of two parts: an outside shell predominantly with $%
3S_{1/2}(F=2)$ atoms, interacting with both lasers, and an inner atomic
cloud mainly pumped to the $3S_{1/2}(F=1)$ ground state. The inner part does
not interact with the lasers and therefore is not subjected to the
limitations of density discussed before. A schematic diagram of the physical
situation is presented in figure 4, where the relative dimensions for both
parts observed are represented.

\begin{figure}[h]
\centering   \includegraphics[width=2in]{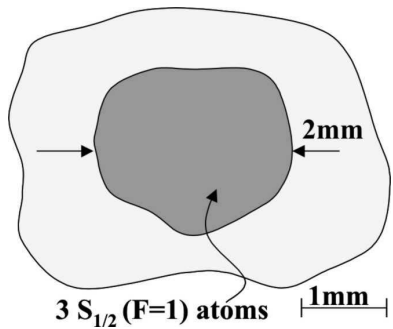}
\caption{ \emph{Schematic representation of the relative atomic distribution
of }$S_{1/2}(F=1)$\emph{\ atoms (dark part in the center) embedded in a
larger trapped cloud.} }
\end{figure}

\bigskip

Using a weak probe beam, which could be tuned resonant with the $F=1$ and $%
F=2$ atomic transition, we have characterized the cloud with respect to both
hyperfine ground state. In order to do that we imaged the probe beam on a
CCD and observed the integrated absorption distribution along the probe beam
path. A sequence of these images can be observed in figures 5 and 6. The
maximum absorption observed of $F=2$\ atoms was less than 50\%, while for
the $F=1$\ atoms was higher than 95\%.\emph{\ }The outer diameter of the $%
3S_{1/2}(F=2)$\ atoms was about $4$\ $mm$ when observe through the
absorption image, the absorption for the $3S_{1/2}(F=1)$\ shows a spatial
distribution within $2$\ $mm$\ of diameter. With the value of absorption and
optical path we can calculate densities which are \ $\thicksim 10^{10}$ $%
cm^{-3}$ for $F=2$ atoms and $\ \thicksim 10^{11}$ $cm^{-3}$ for $F=1$
atoms. This situation characterizes what we have called a 'self dark SPOT',
where the inner part of the cloud is in the dark and the outside works as a
regular MOT. These numbers are in agreement with the conventional dark SPOT
reported in ref. \cite{KDJ}.

We have observed that to obtain the self-induced dark SPOT the conditions on
the repumper detuning and the efficiency of deceleration process are
crucial. The figure 5 shows a sequence of absorption images for the $%
3S_{1/2}(F=1)$ atoms in different conditions of deceleration. As the
magnetic field of the slower magnet is lowered, more atoms emerge in the $%
S_{1/2}(F=1)$ state and the establishment of the SDMOT\textit{\ }is
remarkable. Usually, when the SDMOT is operating, one can see a strong
absorption in a weak $F=1$ probe beam with bare eyes.

\begin{figure}[h]
\centering   \includegraphics[width=3.5in]{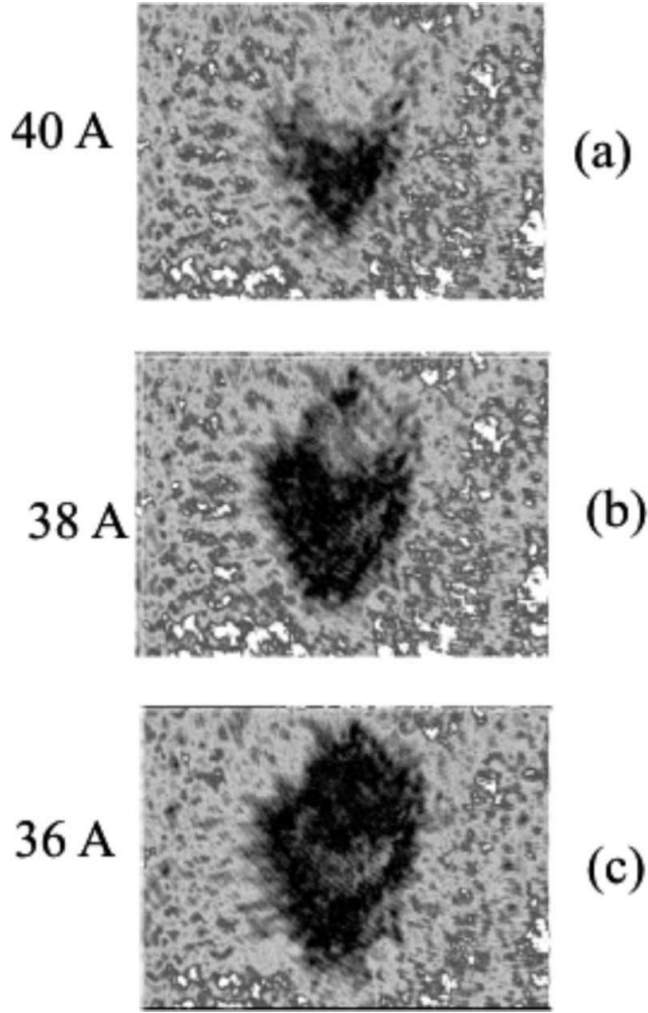}
\caption{ \emph{Sequence of images showing the variation in the absorption
of a weak probe beam for different currents of the slowing solenoid. As the
magnetic field decreases a larger flux of atoms is obtained in the F=1
state, increasing the absorption signal. The darkest parts in figure (c)
represent absorption of approximately 100\% in the probe beam.} }
\end{figure}

However, if the flux of $3S_{1/2}(F=1)$ atoms is not high enough to create
the initial density profile, the subsequent pumping of the atoms to $%
3S_{1/2}(F=1)$ is not reached. It is interesting to note that is not only
the absorption of the repumper that plays a role in the creation of a
population difference across the cloud, but also the decrease in the
absorption of the trap laser, which due to the higher intensity rapidly
pumps the atoms to the lower hyperfine and accelerates the establishment of
the differential population profile.\emph{\ }

\begin{figure}[h]
\centering   \includegraphics[width=3.5in]{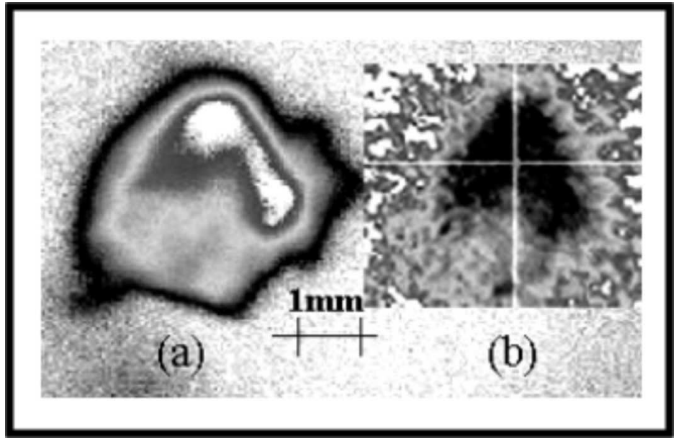}
\caption{ \emph{Comparison between a fluorescence (a) and absorption (b)
images from the same SDMOT. The figure shows the relative size of the "dark
cloud" (}$S_{1/2}(F=1)$\emph{\ atoms) embebed within the whole trapped cloud.%
} }
\end{figure}

Figure 6 shows two images of the trap, where the first one (a) was obtained
by fluorescence and the second (b) by absorption imaging. The absorption
image was obtained with a weak probe laser, tuned to the $F=1\longrightarrow
F^{\prime }=2$. Both images have the same scale and where taken while the
SDMOT was turned on. It is clearly observed that the absorption image of the 
$F=1$ atoms is concentrated in the inner part of the trap and its spatial
distribution is smaller than the whole trap, which is represented in the
fluorescence image. Irregularities on the trap image are mainly due to laser
beam inhomogeneity and certain trap instabilities due to the high number of
atoms.

\section{A simple model for the SDMOT}

In order to understand our observations we present here a very simple
analytical model to show the attenuation of the laser intensity along the
cloud and the accumulation of the $F=1$ state atoms at the central part of
the trap. The idea of the model is simply to point out the main features
presented by the system. Our model considers the four-level system shown in
figure 7. The levels (1) and (2) are respectively the ground states $%
3S_{1/2}(F=1)$ and $3S_{1/2}(F=2)$ while the levels $(2^{\prime })$ and $%
(3^{\prime })$ are the respective $3P_{3/2}$. We consider this four-level
system in the presence of the two laser beams, the trapping and repumper
laser, both represented in the figure 7 respectively by the frequencies $%
\omega _{1}$ and $\omega _{2}$.

\begin{figure}[h]
\centering   \includegraphics[width=4.4in]{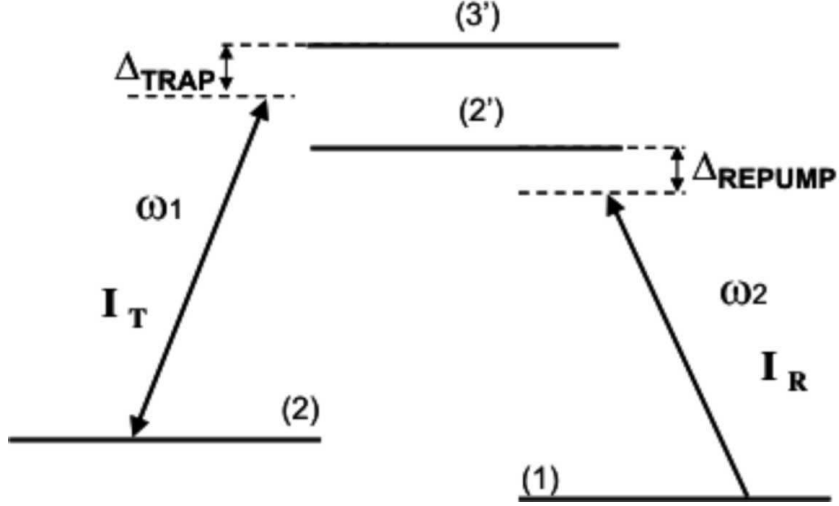}
\caption{ \emph{A four-level diagram used to construct the one-dimensional
model presented. The two lasers are also represented in the diagram.} }
\end{figure}

Using the rate equations for the transitions involved one can determine the
local population ratio as a function of light intensity. In the equations
below, $N_{2}$ is the number of atoms in state 2, while $N_{1}$ is the
number in state 1. The local population ratio on steady state is given by:

\begin{equation}
\frac{N_{2}}{N_{1}}=\frac{2P_{12^{\prime}}-P_{22^{\prime}}}{P_{23^{\prime}}}
\tag{1}
\end{equation}
where 
\begin{equation}
P_{12^{\prime}}(\omega_{2})=\frac{1}{2}\frac{\Omega_{_{12^{\prime}}}^{2}(x)/2%
}{\Delta_{repump}^{2}+(\Gamma/2)^{2}+\Omega_{_{12^{\prime}}}^{2}(x)/2} 
\tag{2}
\end{equation}

\begin{equation}
P_{22^{\prime }}(\omega _{1})=\frac{1}{2}\frac{\Omega _{_{22^{\prime
}}}^{2}(x)/2}{(58+\Delta _{trap})^{2}+(\Gamma /2)^{2}+\Omega _{_{22^{\prime
}}}^{2}(x)/2}  \tag{3}
\end{equation}

\begin{equation}
P_{23^{\prime}}(\omega_{1})=\frac{1}{2}\frac{\Omega_{_{23^{\prime}}}^{2}(x)/2%
}{\Delta_{trap}^{2}+(\Gamma/2)^{2}+\Omega_{_{23^{\prime}}}^{2}(x)/2}  \tag{4}
\end{equation}

\bigskip

The terms $\Omega _{_{12^{\prime }}}(x)=5\sqrt{I_{R}(x)},$ $\ \Omega
_{_{22^{\prime }}}(x)=13\sqrt{I_{T}(x)}$ and $\Omega _{_{23^{\prime }}}=28%
\sqrt{I_{T}(x)}$ are the local Rabi frequencies along a given $\mathit{x}$%
-direction, while $I_{T}(x)$ represents the intensity of trap laser and \ $%
I_{R}(x)$ the repumper laser intensity along the same line. In all the
equations the frequencies are expressed in MHz and the intensities in mW/mm$%
^{2},$ as given by the reference \cite{Farrel}. In our case $\Delta
_{trap}=-10$ MHz and $\Delta _{repump}=-80$ MHz (as represented by the
figures 3 and 7). The calculation, which is a simplified one-dimensional
version of the actual trap, starts considering an atomic cloud with
homogenous distribution of $3S_{1/2}(F=1)$ atoms. As the lasers are turned
on, the strong absorption of the repumper laser produces a population
imbalance as predicted by equation (1). At each position, as the intensities 
$I_{R}(x)$ \ and $I_{T}(x)$\ varies, higher concentration of atoms in $F=1$
appears. The final result, predicted by the rate equation model, is the
establishment of a population and laser intensity profiles across the atomic
cloud. At steady state the one-dimensional profile for atomic population and
laser intensity are represented in figure 8.

\begin{figure}[tbh]
\centering   \includegraphics[width=3.5in]{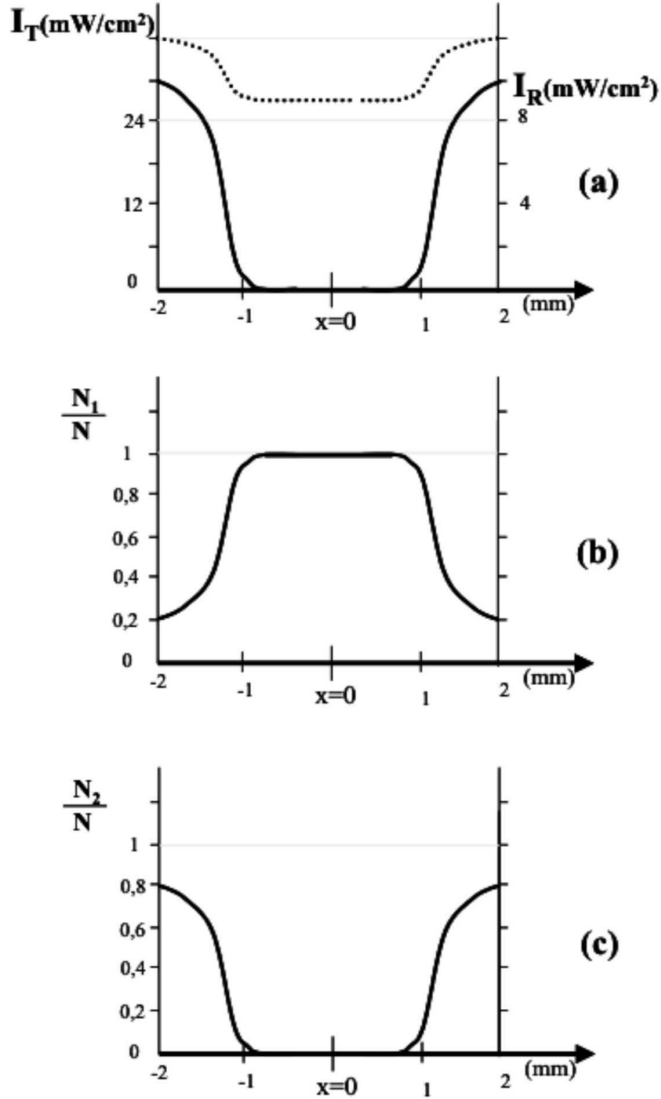}
\caption{ \emph{\ Representation of the numerical results from the
one-dimensional model. The figure shows the intensity profile of both lasers
(a), population in the }$S_{1/2}(F=1)$\emph{\ state (b), and the population
in\ }$3S_{1/2}(F=2)$\emph{\ (c), after reaching the steady state conditions.
In the figure (a) the dash line indicates the trap laser intensity while the
continuous line indicates the repumper laser.} }
\end{figure}

According to the figure 8(a) we see that the intensity of the trapping laser
decreases only modestly, while the repumper is fully attenuated as it
penetrates in atomic cloud. We also see that eventhough the population in $%
F=1$\ is initially small at the outer part of the trapped cloud; it
increases rapidly as the repumper laser is absorbed. The population in $F=2$%
\ is complementary because in our model the local number obeys $%
N_{1}+N_{2}=N.$\ The importance of this model resides in showing the
establishment of a population profile, which is qualitatively in concordance
with our observations.

\section{Conclusions}

We have observed a new kind of dark SPOT, which is naturally produced by
loading a standard MOT from an intense flux of slowed atoms in the $%
3S_{1/2}(F=1)$\ state, and by adjusting the intensity and frequency of the
repumper laser. Using the system described here it was possible to create a
"dark MOT ", without spatially separating the $F=1$ and $F=2$ laser beams.
The shadow in the $F=1$\ light was created\ by the own absorption of the
atoms, due to an accumulation of \ \ "dark state " atoms in the center of
the trap. Because the resemblance of this process with the work of Ketterle
et al \cite{KDJ}, we named this new trap as self-induced dark MOT (SDMOT).

"Dark MOTs " of this kind, where atoms stay predominantly in a "dark"
hyperfine level, not interacting with the trapping light, are an important
to overcome density limitations of the standard MOT. We verified that the
SDMOT works as good as the usual dark SPOT configuration, with the advantage
that since it does not require spatially separated beams for trap and
repumper, it can be simply operated with a single electro-optical modulator
(EOM) to produce the repumping light. This simplifies the experimental setup
also because it does not require special alignments to image the blocking
disk, used as a physical obstacle in the repumper beam. However, the
situation we have presented here depends upon a dense flux of atoms in
conditions to be trapped, as well as some special tuning conditions for the
repumping light. Some preliminary results also show that the temperature of
those sample might be lower than the usual dark SPOT, but to understand such
effect a more comprehensive theoretical model will be required, as well as
other experimental investigations.

\section{Acknowledgments}

This work was developed at Center for Research in Optics and Photonics,
CePOF, and it was supported by FAPESP, though the program CEPID. It also had
support from the Brazilian national funding agency, CNPq.

\end{document}